\documentclass[a4paper,english,12pt]{article}
\usepackage[T2A]{fontenc}
\usepackage{hyperref,booktabs,graphicx,cite,enumerate,amsmath,amssymb} 
\usepackage{paralist,tikz,pgfplots,multirow,lscape,multicol,longtable}
\usepackage{rotating,libertine}
\usepackage[right=25mm,left=30mm,top=10mm]{geometry}

\pgfplotsset{compat=1.18}

\title{Query2Diagram: Answering Developer Queries with UML Diagrams\footnote{Submitted to the \href{https://link.springer.com/journal/10958}{Journal of Mathematical Sciences}.}}
\date{}

\author{
  Oleg Baryshnikov\\
  HSE University\\
  \texttt{baryshnikovod@gmail.com}
  \and
  Anton M.~Alekseev\\
  St. Petersburg Department of Steklov Mathematical Institute, RAS,\\
  St. Petersburg State University\\
  \texttt{anton.m.alexeyev@gmail.com}
  \and 
  Sergey~I.~Nikolenko\\
  St. Petersburg Department of Steklov Mathematical Institute, RAS,\\
  St. Petersburg State University\\
  \texttt{sergey@logic.pdmi.ras.ru}
}
 
\begin{document}
\maketitle     
\begin{abstract}
Software documentation frequently becomes outdated or fails to exist entirely, yet developers need focused views of their codebase to understand complex systems. While automated reverse engineering tools can generate UML diagrams from code, they produce overwhelming detail without considering developer intent. We introduce \emph{query-driven UML diagram generation}, where LLMs create diagrams that directly answer natural language questions about code. Unlike existing methods, our approach produces semantically focused diagrams containing only relevant elements with contextual descriptions. We fine-tune Qwen2.5-Coder-14B on a curated dataset of code files, developer queries, and corresponding diagram representations in a structured JSON format, evaluating with both automatic detection of structural defects and human assessment of semantic relevance. Results demonstrate that fine-tuning on a modest amount of manually corrected data yields dramatic improvements: our best model achieves the highest F1 scores while reducing defect rates below state-of-the-art LLMs, generating diagrams that are both structurally sound and semantically faithful to developer queries. Thus, we establish the feasibility of using LLMs for scalable contextual, on-demand documentation generation. We make our code and dataset publicly available at \url{https://github.com/i-need-a-pencil/query2diagram}.

\vspace{2mm}

\end{abstract}

\section{Introduction}\label{sec:intro}

Software documentation, particularly UML diagrams, plays a crucial role in system comprehension and maintenance. Studies confirm that UML usage improves code quality and modularity, reduces defects, and increases developer productivity~\cite{arisholm2006impact,Nugroho2009EvaluatingTI,nugroho2014impact,nugroho2008survey}. However, manually creating and, crucially, \emph{maintaining} up-to-date diagrams requires significant effort, leading to a common problem: documentation that either does not exist or diverges from the actual implementation~\cite{hebig2016quest}.

In the absence of up-to-date UML documentation, diagrams can be generated with \emph{automated reverse-engineering} (RE) tools. However, the resulting UML diagrams often contain overwhelming detail that hinders comprehension; studies and developer surveys show that these diagrams have to be severely simplified for clarity~\cite{osman2012uml,osman2013uml}. Some approaches propose interactive filtering mechanisms~\cite{egyed2002automated}, but typically ignore semantic context and user intent. This creates a gap: developers need focused, contextual views of their codebase, not exhaustive structural dumps.

\begin{figure}[!t]
    \centering
    \includegraphics[width=0.85\linewidth]{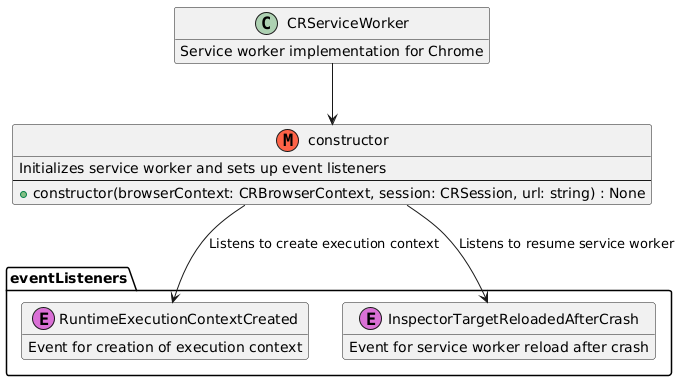}
    \caption{A sample response to the user query ``Map out the event listeners set up in the CRServiceWorker constructor and their corresponding  actions'' based on the TypeScript file \href{https://github.com/microsoft/playwright/blob/main/packages/playwright-core/src/server/chromium/crServiceWorker.ts}{microsoft/playwright/... /crServiceWorker.ts}.}
    \label{fig:queryexample}
\end{figure}

Recent advances in NLP, specifically large language models (LLMs), offer a promising solution. LLMs excel at understanding both code and natural language, making them uniquely suited to bridge the gap between developer queries and visual representations. Although LLM applications in software engineering are expanding rapidly~\cite{zhang2023survey,lomshakov2024large}, including tools such as GitHub Copilot~\cite{Copilot} and Cursor~\cite{Cursor}, their potential for query-driven diagram generation remains unexplored.

Motivated by this gap, we present a novel approach: using LLMs to generate RE-UML diagrams that \emph{directly answer developer queries} about code. Unlike traditional RE tools or recent LLM-based approaches that generate complete diagrams, our method produces \emph{query-focused} diagrams that include only relevant elements with contextual descriptions. Fig.~\ref{fig:queryexample} illustrates this: given a query about event listeners in a TypeScript file, the system generates a targeted diagram showing only the relevant components and their relationships.

Our approach offers several key advantages: it (1) eliminates the need for pre-existing diagrams, (2) provides direct answers to developer questions through visual representation, (3) includes explanatory descriptions for all elements, and (4) supports multiple programming languages through a single model. We address two research questions:

\begin{itemize}    
    \item[\textbf{RQ1}] Can LLMs generate relevant, comprehensible graph-like structures (UML diagrams) when provided with code files and high-level queries about design patterns, component interactions, or architectural concerns?
    \item[\textbf{RQ2}] Can training on synthetic and curated data effectively control the quality properties of LLM-generated diagrams?
\end{itemize}

In this work, we fine-tune Qwen2.5-Coder-14B on a carefully curated dataset of code-query-diagram triples, introducing a JSON-based intermediate representation that reduces syntax errors while enabling structured generation. Our dual evaluation strategy combines automatic defect analysis with human relevance assessment. Results demonstrate that fine-tuning on manually corrected data yields diagrams that are both structurally sound and semantically relevant, achieving the best F1 scores while dramatically reducing defect rates.

Our contributions include: (1) a new task formulation for query-driven UML generation from code, (2) a practical method using fine-tuned LLMs with structured output generation, (3) a comprehensive evaluation framework combining structural and semantic metrics, and (4) empirical evidence that even small amounts of high-quality training data can significantly improve diagram quality. We release our code and dataset at \url{https://github.com/i-need-a-pencil/query2diagram}.

In the following, Section~\ref{sec:relatedwork} reviews related work on diagram generation, Section~\ref{sec:method} provides a description of our approach, Section~\ref{sec:evaluation} presents evaluation results and discussions, and Section~\ref{sec:conclusion} concludes the article.

\section{Related Work}\label{sec:relatedwork}

Recent studies document both the demand for and the friction in keeping diagrams useful in practice: informal diagrams dominate real projects yet are rarely updated~\cite{jongeling2025informal}; API–scale structure benefits from interactive, automatically generated visualizations~\cite{stepanek2025helveg}; and large ecosystems require continuously refreshed, machine-generated views to track versions, dependencies, and CVEs~\cite{kana2025automated}. Reverse-engineered models also drive downstream tasks beyond comprehension, such as early defect prediction from UML metrics~\cite{battulga2025metric} and guided system migration~\cite{khiati2025wa2ma}. These findings jointly motivate \emph{focused}, on-demand diagramming from code, rather than one-shot, complete diagrams that quickly drift from developers' information needs.

\textbf{Traditional UML Generation.}
Classical reverse engineering (RE) approaches for UML generation fall into three categories. \emph{Static analysis} tools~\cite{tonella2001reverse,sutton2007recovering} extract structural relationships without code execution; they effectively detect inheritance but struggle with runtime behaviors. 
\emph{Dynamic analysis} tools such as Caffeine~\cite{gueheneuc2002no} capture runtime behavior and can discover specific object relationships, but require executable code and sufficient coverage. 
\emph{Hybrid approaches} like Ptidej~\cite{gueheneuc2004reverse}, which enhances a static model with dynamic traces, combine both but demand complex heuristics and manual configuration. {Beyond comprehension, traditional RE feeds downstream uses (e.g., migration~\cite{khiati2025wa2ma} or defect prediction from UML metrics~\cite{battulga2025metric}), yet the shared limitation remains: they generate \emph{complete} structural views without considering user intent or information needs.}

\textbf{LLM-Based UML Generation.}
Recent research explores large language models (LLMs) for UML generation from both source code and natural language artifacts. 
\cite{shehata2024creating} used GPT-4~\cite{achiam2023gpt} to generate class diagrams directly from codebases, reporting frequent syntax errors and missing elements.  
{\cite{siala2025using} embedded an LLM into a model-driven reverse-engineering (MDRE) pipeline to extract class diagrams from Java programs, while ~\cite{siala2025leveraging} fine-tuned Mistral-7B on AgileUML-derived Java/Python $\leftrightarrow$ UML/OCL pairs, achieving high precision and recall; ~\cite{siala2025towards} targeted explicit OCL generation from code, substantially outperforming rule-based baselines; and ~\cite{boronat2025mdrellm} combined LLMs with MDRE, introducing diagram-granularity levels (CLASS, COARSE, FINE) to control abstraction.}
{\cite{amalfitano2025automated} evaluated GPT-4o, Gemini 1.5 Pro, Claude 3.5 Sonnet, and Mistral Large on architecture recovery, showing that self-reflection prompts reduce hallucinations and omissions.} 

{Beyond code, several studies use LLMs to translate textual specifications into UML:~\cite{camara2023assessment,debari2024,ferrari2024model,jahan2024} focus on requirements-to-model generation, while~\cite{babaalla2025llm} applies a hybrid NLP–DSL–MDA pipeline to produce UML and executable code.}

Most related to our focus on intent conditioning,~\cite{ben2024software} showed that guiding diagram construction through natural-language queries can improve productivity and coherence---but their system assumes a pre-existing UML model rather than generating one from code.

Overall, existing LLM approaches either generate complete diagrams from source or derive them from textual specifications; none support \emph{query-driven} diagram generation that filters and abstracts directly \emph{from code} according to developer information needs.

\textbf{Quality Assessment.}
UML diagram quality is typically evaluated via \emph{syntactic validation}~\cite{unhelkar2005verification} and \emph{metric-based analysis} of structural properties~\cite{genero2005survey}, or by element-level precision/recall/F1 against a reference model. Code $\rightarrow$ UML/OCL work reports such structural metrics---for example, LLM-in-the-loop and fine-tuned pipelines benchmark element extraction against rule-generated ground truth~\cite{siala2025using,siala2025leveraging,siala2025towards}, while~\cite{boronat2025mdrellm} compares LLM outputs with a strong procedural baseline using granularity-aware metrics (CLASS/\allowbreak COARSE/\allowbreak FINE) and WCC. 
{Architecture-oriented recovery additionally analyzes error taxonomies (missing/mistake/hallucination) and the effect of self-reflection prompts~\cite{amalfitano2025automated}.} 

{By contrast, NL$\rightarrow$UML studies (from specifications rather than code) rely largely on \emph{manual annotation} and rubric-based judgments of syntactic/semantic quality, frequently noting syntax errors or missing elements in raw LLM outputs~\cite{shehata2024creating,jahan2024,ferrari2024model}.} 
Across these strands, prevailing evaluations focus on \emph{structural correctness}, not \emph{task relevance}: they rarely assess whether a generated diagram actually answers a developer's \emph{query}.
{Practitioner evidence that informal diagrams are ubiquitous yet seldom maintained~\cite{jongeling2025informal} and ecosystem-scale visualization needs for continuously updated, actionable overviews~\cite{kana2025automated} further suggest prioritizing \emph{focused}, on-demand views.} 
{This motivates relevance-oriented metrics that evaluate whether a diagram produced \emph{from code} satisfies a specific information need.}

\textbf{Positioning our work.}
The literature above separates into (i) traditional analysis-based tools that output exhaustive diagrams and (ii) LLM-driven methods that either generate exhaustive diagrams from code~\cite{siala2025using,siala2025leveraging} or operate from specifications~\cite{babaalla2025llm}, with granularity controls but no intent conditioning~\cite{boronat2025mdrellm}. We introduce a different paradigm: \emph{query-driven diagram generation from code}, producing \emph{focused, contextual} views guided by developer intent. Unlike classical tools, we do not emit ``everything,'' and unlike specification-driven or architecture-only studies~\cite{amalfitano2025automated}, our inputs are the source files themselves. Our evaluation complements syntax and element-matching with \emph{relevance-oriented} metrics that ask whether the generated view answers the user's query. To the best of our knowledge, no prior work combines code input with \emph{query-conditioned} selection and abstraction for UML generation.

Concretely, we operationalize queries over code by retrieving intent-aligned fragments and constraints, prompting an LLM to synthesize a minimal UML view, and validating the view against both structural consistency and query relevance.

\section{Method}\label{sec:method}
Below we describe our workflow consisting of three stages: \emph{data collection}, \emph{model adaptation}, and \emph{diagram generation}. Each stage is deliberately lightweight so that new code bases or languages can be added with minimal manual effort.

\textbf{Data curation: code selection.}
Using the public GitHub metadata dump by~\cite{elmerscode}, we collected the 150 most‑starred repositories with permissive licences (MIT, MIT‑0, or Apache‑2.0), covering twelve mainstream languages (C, C++, Java, Python, JavaScript, TypeScript, Rust, PHP, C\#, Scala, Kotlin, Go); multiple languages serve as an additional test for the multi-lingual capabilities of the resulting model. From each repository, we filtered source files in the range of 3K--15K characters, discarded near‑duplicates (via Jaccard similarity of simple unigram representations) and non‑ASCII files, and stratified the final dataset into a \emph{train/val/test} split with 88/12/24 files respectively, keeping all languages represented.

\textbf{Data curation: query synthesis.}
We know of no public benchmark currently linking \emph{questions about code} to UML diagrams;~\cite{hoque2022chart} notes that in the absence of human-written questions, one can use templates or NLP-based generation methods~\cite{ghaleb2018program}. We generated user queries with two open LLMs, DeepSeek-R1-Distill-Llama-70B~\cite{deepseekai2025deepseekr1incentivizingreasoningcapability} and QwQ-32B-Preview~\cite{qwq-32b-preview} (50\% each). Prompt templates detailed in Fig.~\ref{tab:query-generation} encourage questions about architecture, API usage, or design patterns that can be answered by a single diagram. Each query is also assigned a detail level: \textit{minimal}, \textit{moderate}, or \textit{full}.

\begin{figure}[ht]
    \centering
    \scriptsize
    \begin{tabular}{|p{\textwidth}|}
        \hline
        \begin{verbatim}
1. The final output must be a JSON array of strings representing questions, 
   enclosed within `<candidates>` and `<final_output>` XML tags. 
   You must strictly follow this template for the final answer: 
   `<candidates>["question1", ..., "questionN"]</candidates>
   <final_output>["question1", ..., "questionK"]</final_output>`.
   During processing, you may represent questions in any format.  
2. Questions must focus on aspects of the provided code, such as project structure,
   component interactions, imported modules, code behavior, design patterns, 
   code design principles, external API usage, deployment strategies, 
   or the potential integration of new components into the code.  
3. Each question will be used as a user query for another LLM.
   Therefore, each generated question should either ask the LLM for information 
   or instruct it to perform a task.
4. The expected answer for each question must be some kind of diagram
   (e.g., PlantUML or Mermaid.js). 
   Ensure each question can be effectively answered with a diagrammatic representation
   rather than just text.  
5. Do not mention diagrams explicitly; instead, use similar words such as structure,
   relationships, interactions and so on. 
6. You must generate 10 candidate questions within the `<candidates>` tags.
7. Filter or combine candidate questions to create a final selection 
   of 0 to 3 high-quality questions within the `<final_output>` tags. 
   Prioritize question quality over quantity.
   Try to cover different topics in the final selection.  
8. The LLM answering the questions will only have access to the content 
   of the provided code file. Therefore, ensure all generated questions 
   can be answered solely based on the information within that file, 
   without requiring external context or knowledge beyond the given code.  
9. Each question must be straightforward and as short as possible. 
   Since each question is intended to be answered by a single diagram,
   avoid combining multiple questions into one.  
10. Each question must be unique. 
    Avoid creating duplicate or near-duplicate questions, even with slight
    variations in wording.
11. If you cannot identify any questions that meet all the specified criteria, 
    it is perfectly acceptable to return an empty list 
    within the `<final_output>` tags. 
    If a question does not fully meet all requirements (or if you're not 100% sure),
    you should remove it.

Generate a list of questions based on the following code file:
```
{code}
```
    \end{verbatim}
    \\ \hline 
    \end{tabular}
    \caption{Prompt template used to generate user queries (sampling, temperature 0.6, top\_p~0.9).}
    \label{tab:query-generation}
\end{figure}

\textbf{Data curation: diagram construction.}
While public documentation, e.g., in the Lindholmen dataset~\cite{hebig2016quest} contains UML diagram images from open source repositories, these diagrams often span multiple files or entire projects, making them GPU-intensive and, more importantly, unaligned with specific user queries; moreover, most diagrams are available only as images, complicating conversion to structured formats.

\begin{figure}[ht]
    \centering
    \scriptsize
    \begin{tabular}{|p{\textwidth}|}
        \hline
        \begin{verbatim}
1. You should generate Python list of nodes, list of edges and list of packages. 
   Each element is presented by JSON.
2. Make sure the final diagram is comprehensible: it should be readable,
   understandable by user.
3. For each node write a short description.
4. The diagram should contain all important nodes and edges to code understanding.
5. You can omit some of the standard, boilerplate-code steps.
6. You can add conceptual entities to diagram 
   (high-level concepts that are not functions or classes).
7. Try to group related nodes (including those you introduce) using package elements,
   where possible, do not use classes for this purpose.
8. Package elements can be nested.
9. Do not add fake classes to aggregate code entities, use packages for this purpose. 
   All fields and methods should be presented as is in code.
10. JSON template for node: {
    "type": Literal["class", "variable", "function", "entity", "method", "field"], # a 
    type of node from the predefined types; use the most appropriate
    "name": str, # the actual name of the node to show on the diagram
    "node_id": str, # id or unique name of the node; use the actual name of the node if 
    possible
    "description": str, # a short description of node
    "visibility": Literal["private", "protected", "package private", "public"], # an 
    access modifier of the node from the predefined types; use the most appropriate
    "return_type": Optional[str], # a return type for functions and methods or current 
    node type for fields and variables; can be skipped for other types of nodes
    "params": Optional[str], # parameters of functions and methods (as in brackets); 
    can be skipped for other types of nodes
    "source_class_id": Optional[str], # node id of the source class for fields and 
    methods; can be skipped for other types of nodes
}
11. JSON template for edge: {
    "node_id_from": str, # id of the node where the edge starts
    "node_id_to": str, # id of the node where the edge ends
    "description": Optional[str], # a short description of the edge; can be None
}
12. JSON template for packages: {
    "package_id": str, # id or unique name of the package; it will be shown on the 
    diagram
    "children": List[str], # list of ids of nodes to include in the package or names of 
    nested packages
    "description": Optional[str], # a short description of the package; can be None
}

\end{verbatim}
\\ \hline 
\end{tabular}
\end{figure}

\clearpage

\begin{figure}[ht]
    \centering
    \scriptsize
    \begin{tabular}{|p{\textwidth}|}
        \hline
        \begin{verbatim}
13. JSON template for graph: {
    "nodes": [{node1_JSON}, ..., {nodeN_JSON}],
    "edges": [{edge1_JSON}, ..., {edgeN_JSON}],
    "packages": [{package1_JSON}, ..., {packageN_JSON}]
}
14. Prefer to use camelCase, snake_case, PascalCase for names, 
    do not use spaces in names.
15. Package names can not be in edges list, use them only in packages.
16. Do not write any explanations, just write expected output.
17. You should generate three versions of the `Graph template`. 
    The first one, called the "minimal version," should include 
    only the essential nodes and logic, removing all unnecessary elements. 
    The second one, the "medium version," should contain all important nodes 
    along with some additional details. 
    The third one, the "full version," should incorporate every possible node 
    and all possible edges relevant to query.
18. After generating all three graph templates, you should provide the shortest 
    possible "text answer" to the user's query using the generated diagrams.
19. Expected output template: {
    "minimal_version": {graph_JSON},
    "medium_version": {graph_JSON},
    "full_version": {graph_JSON},
    text_answer: str,
}

Your task is write nodes, edges and packages to build 
a diagram in different formats (PlantUML, mermaid-js, ...)
for the following file in project:
```
{code}
```
You need to generate a diagram for the following user query:
\"{query}\"

\end{verbatim}
\\ \hline 
\end{tabular}
\caption{Prompt template used to generate diagrams with base models (greedy).}
\label{tab:diagram-by-query}
\end{figure}

Thus, we used a hybrid approach: first, queries were answered by \emph{Claude 3.5 Sonnet}, which yielded the highest JSON validity in preliminary tests\footnote{We compared with the best models available in early 2025: GPT-4o, o1-preview, o3, QwQ-32B-Preview, DeepSeek-R1, DeepSeek-R1-Distill-Llama-70B, DeepSeek-R1-Distill-Qwen-32B, Qwen2.5-Coder-7B-Instruct, Qwen2.5-Coder-14B-Instruct, and Qwen2.5-Coder-32B-Instruct.}; the prompt template is given in the Table~\ref{tab:diagram-by-query}. The output is a \emph{format‑agnostic JSON graph} with lists of \emph{nodes}, \emph{edges}, and (nested) \emph{packages}. Six element types are allowed: \emph{class}, \emph{method}, \emph{field}, \emph{function}, \emph{variable}, and \emph{abstract entity}, with every element and connections between them containing a brief description (see full schema in Table~\ref{tab:jsondesc}). We initially produced $264$ training and $36$ validation graphs.

\begin{table}[!htbp]
\renewcommand{\arraystretch}{1.7}
\small
\centering

\label{tab:jsondesc}
    \begin{tabular}{p{2.0cm}|p{2.4cm}|p{1.9cm}|p{2.7cm}}
    \toprule
    \textbf{Component} & \textbf{Field Name} & \textbf{Type} & \textbf{Description} \\ \midrule
    Output & nodes & List[dict] & List of nodes \\
    \cline{2-4}
    & edges & List[dict] & List of edges \\
    \cline{2-4}
    & packages & List[dict] & List of packages \\
    \cline{1-4}
    Node & type & str & Node type \\
    \cline{2-4}
    & name & str & Display name \\
    \cline{2-4}
    & node\_id & str & Unique ID \\
    \cline{2-4}
    & description & str & Short description \\
    \cline{2-4}
    & visibility & str & Access modifier \\
    \cline{2-4}
    & return\_type & Optional[str] & Return or data type \\
    \cline{2-4}
    & params & Optional[str] & Method/function parameters \\
    \cline{2-4}
    & source\_class\_id & Optional[str] & ID of the parent class \\
    \cline{1-4}
    Edge & node\_id\_from & str & Source node ID \\
    \cline{2-4}
    & node\_id\_to & str & Target node ID \\
    \cline{2-4}
    & description & Optional[str] & Edge label \\
    \cline{1-4}
    Package & package\_id & str & Package name \\
    \cline{2-4}
    & children & List[str] & Included node or sub-package IDs \\
    \cline{2-4}
    & description & Optional[str] & Package description \\
    \bottomrule
    \end{tabular}
\caption{JSON graph format and component descriptions.}
\end{table}

\begin{figure}[!t]
    \centering
    \includegraphics[width=0.77\linewidth]{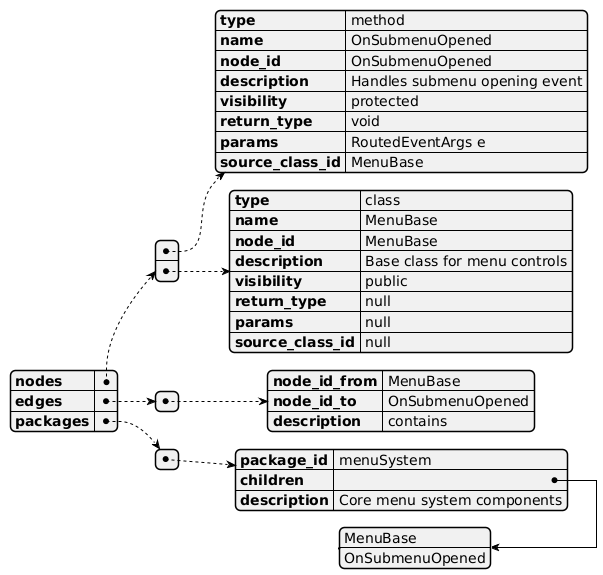}
    \includegraphics[width=0.58\linewidth]{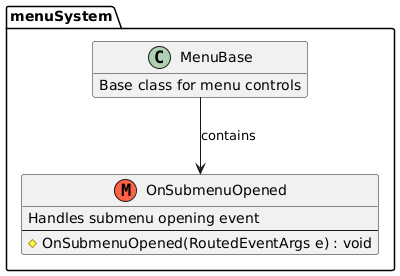}
    \caption{A sample JSON graph and visualization rendered with PlantUML's toolkit.}
    \label{fig:jsonexample}
\end{figure}

Then, due to frequent minor and severe defects (see a list in the Table~\ref{tab:defects}), all graphs were manually reviewed and corrected until all defects were fixed. Graphs with irreparable defects, e.g. with only a single node, were discarded. The resulting JSON graphs can be converted to PlantUML or Mermaid for visualization; we have experimented with generating diagrams directly in markup languages but encountered frequent syntax errors, consistent with the observations by~\cite{shehata2024creating}. Fig.~\ref{fig:jsonexample} shows a sample graph and a rendered diagram.

\textbf{Model adaptation and generation: fine-tuning.}
As the backbone, we chose \emph{Qwen‑2.5‑Coder‑14B‑Instruct} due to its strong zero‑shot coding performance within a single‑GPU budget (NVIDIA Tesla V100, 32GB VRAM). To fine‑tune the 14B model, we used {QLoRA}~\cite{dettmers2023qlora} which quantizes $W_0$ to a 4‑bit NF4 representation and full precision adapters, training with supervised cross‑entropy loss on the (query, diagram) pairs. We used LLaMA-Factory~\cite{zheng2024llamafactory} and incorporated Unsloth~\cite{unsloth} optimizations including L2 regularization, dropout, cosine learning rate scheduling, gradient accumulation, checkpointing, and FP16 precision. Hyperparameters are listed in Table~\ref{tab:hypers}.

\begin{table}[!t]
    \centering

    \label{tab:hypers}
    \begin{tabular}{rc|rc}
    \toprule
    \textbf{Hyperparameter} & \textbf{Value} & \textbf{Hyperparameter} & \textbf{Value} \\ 
    \midrule
    Learning rate & $3\mathrm{e}{-4}$ & Dropout & $0.05$ \\
    L2 regularization & $1\mathrm{e}{-7}$ & Batch size & $2$ \\
    LoRA rank & $64$ & Gradient accumulation & $32$ \\
    LoRA $\alpha$ & $64$ & Training steps & $24$ \\
    \bottomrule
    \end{tabular}
    \caption{Training hyperparameters.}
\end{table}

\textbf{Model adaptation and generation: inference.}
At inference, we employ \emph{vLLM}~\cite{kwon2023efficient} for fast token streaming and \emph{Outlines}~\cite{willard2023efficient} to impose the JSON schema as a constrained grammar, guaranteeing syntactically valid graphs even for small models. The same prompt template (see Fig~\ref{tab:diagrams-finetuned-prompt}) was used for both fine-tuning and decoding.

\begin{figure}[ht]
    \centering\small
    \begin{tabular}{|p{\textwidth}|}
        \hline
    \begin{verbatim}
Your task is to generate json with "nodes", "edges" and "packages" 
for a diagram.
The diagram must answer to the user query using the code.

<code>
{code} </code>
<query>
{query} [{version} version]

\end{verbatim}
    \\ \hline 
    \end{tabular}
    \caption{Prompt template used to generate diagrams with fine-tuned models (greedy).}
    \label{tab:diagrams-finetuned-prompt}
\end{figure}

\begin{table}[!t]
    \centering

    \label{tab:relevancy}
    \begin{tabular}{p{2.5cm}p{7.6cm}}
    \toprule
       \textbf{Metric} & \textbf{Description} \\ \midrule
       \textbf{Suffi\-ciency} & \# of nodes that \emph{must} be in the response to answer the query \\ 
       \hline
       \textbf{Comple\-teness} & \# of nodes that are not required to be present in the response, but helpful for understanding of the answer \\
       \hline
       \textbf{Halluci\-nations} & \# of nodes that cannot be derived from the code, e.g., ``invented'' variables and classes \\ 
       \hline
       \textbf{Verbo\-sity} & \# of nodes irrelevant to the query in a diagram \\ 
    \bottomrule
    \end{tabular}
    \caption{Metrics designed to evaluate whether each node is relevant to the user's information need.}
\end{table}

\section{Evaluation criteria}\label{sec:evaluationcriteria}

We assess our system from two complementary angles: \emph{structural soundness} of the graphs and their \emph{semantic adequacy} with respect to the query.

\begin{table}[!t]
    \centering\setlength{\tabcolsep}{3pt}

    \label{tab:labeling}
    \begin{tabular}{lcccc}
    \toprule
       \textbf{Model} & $\mathbf{Su}\uparrow$& $\mathbf{Co}\uparrow$& $\mathbf{Ha}\downarrow$& $\mathbf{Ve}\downarrow$ \\ \midrule
       GPT-4o & $181$ & $51$ & $\mathbf{0}$ & $\mathbf{13}$ \\
       Claude-3.5-Sonnet & $234$ & $76$ & $1$ & $14$ \\ 
       Qwen2.5-Coder-14B & $252$ & $\mathbf{159}$ & $19$ & $115$ \\
       $\quad$+SFT: Claude Synth & $\mathbf{258}$ & $150$ & $10$ & $61$ \\ 
       $\quad$+SFT: Fixed Claude Synth & $\mathbf{258}$ & $145$ & $4$ & $62$ \\ 
    \bottomrule
    \end{tabular}

    \caption{Number of nodes by relevance classes. \textbf{Su}~--- Sufficiency, \textbf{Co}~--- Completeness, \textbf{Ha}~--- Hallucinations, \textbf{Ve}~--- Verbosity.}
\end{table}

\textbf{Automatic defect analysis.}
First, a Python checker scans every JSON graph for 19 defect patterns grouped into \emph{minor} (stylistic issues, suspicious structure, missing expected connections etc.), \emph{severe} (when components have to be removed or modified), and \emph{unacceptable} (unrenderable) categories (for a list of defects see Appendix, Table~\ref{tab:defects}).  We report (i) \emph{micro} defects per
node, (ii) \emph{macro} defects per node averaged
per diagram, and (iii) the mean defects per diagram.

\textbf{Human relevance annotation.}
Second, to assess the actual relevance of the diagram, two experts independently classify each node into \textsc{Sufficiency}, \textsc{Completeness}, \textsc{Hallucination}, or \textsc{Verbosity} categories (see Table~\ref{tab:relevancy}). Based on confusion counts, we compute standard classification metrics: precision, recall, and F$_1$, plus their \emph{hard} versions
that treat only \textsc{Sufficiency} as true positives (see a full description in Table~\ref{tab:metrics_formulae}).

\textbf{Metrics.}
It is difficult to provide a ``gold set'' of nodes that must be present in a diagram for a given query and code, as diagrams can be built with different approaches; e.g., some models group configuration parameters, with each group responsible for a specific routine, thereby omitting dozens of hashmap fields. 

Still, measuring recall is clearly important, so we used a compromise: the number of false negatives (FN) is estimated as the difference between the maximum number of relevant nodes found across all models for a given query and the number of true positives for a specific model. Note that these metrics (marked with an asterisk $\star$ in Table~\ref{tab:metrics_formulae}) differ from standard definitions. Since $\mathrm{FN}$ and $\mathrm{FN_{hard}}$ represent lower bounds, the corresponding $\mathrm{Recall}$, $\mathrm{F1}$, $\mathrm{Recall_{hard}}$, and $\mathrm{F1_{hard}}$ metrics are upper-bound estimates.

\begin{table}[!t]
     \small\centering\setlength{\tabcolsep}{2pt}

    \label{tab:metrics_formulae}
    \begin{tabular}{p{0.4\linewidth}|p{0.55\linewidth}}
    \toprule
       \textbf{Metric} & \textbf{Formula} \\ \midrule
       $\mathrm{TP}$ & $\mathrm{|Su| + |Co| }$ \\ 
       \hline
       $\mathrm{FP}$ & $\mathrm{|Ha| + |Ve|}$ \\
       \hline
       $\mathrm{FN}\star$ & $\mathrm{\mathbf{max}(|Su|)+\mathbf{max}(|Co|)-TP}$ \\ 
       \hline
       $\mathrm{TP_{hard}}$ & $\mathrm{|Su|}$ \\ 
       \hline
       $\mathrm{FN_{hard}}\star$ & $\mathrm{\mathbf{max}(|Su|)-TP_{hard}}$ \\ 
       \hline
       $\mathrm{Precision}$ & $\mathrm{{TP}/{(TP+FP)}}$ \\ 
       \hline
       $\mathrm{Recall}\star$ & $\mathrm{{TP}/{(TP+FN)}}$ \\ 
       \hline
       $\mathrm{F1}\star$ & $\mathrm{{2*TP}/{(2*TP+FP+FN)}}$ \\ 
       \hline
       $\mathrm{Precision_{hard}}$ & $\mathrm{{TP_{hard}}/{(TP_{hard}+FP)}}$ \\ 
       \hline
       $\mathrm{Recall_{hard}}\star$ & $\mathrm{{TP_{hard}}/{(TP_{hard}+FN_{hard})}}$ \\ 
       \hline
       $\mathrm{F1_{hard}}\star$ & $\mathrm{{2*TP_{hard}}/{(2*TP_{hard}+FP+FN_{hard})}}$ \\ 
    \bottomrule
    \end{tabular}

    \caption{Metrics. $\mathrm{|Su|}$---``Sufficiency'' nodes count, $\mathrm{|Co|}$---``Completeness'', $\mathrm{|Ha|}$---``Hallucinations'', $\mathrm{|Ve|}$---``Verbosity''.}
\end{table}

\begin{table}[!t]
    \small\centering\setlength{\tabcolsep}{2pt}

    \label{tab:defects}
    \begin{tabular}{p{0.47\linewidth}|p{0.47\linewidth}}
    \toprule
        \textbf{Minor Defects} & \textbf{Severe Defects}\\ 
        \hline
        Spaces in node names & Non-unique packages ids \\
        \hline
        Spaces in package ids & Non-unique nodes ids \\
        \hline
        Single node & Edges from/to non-valid node ids \\
        \hline
        No edges & Methods/fields have empty source class id \\
        \hline
        Edge to itself & Packages without nodes \\
        \hline
        Repeated edges & Child in multiple packages \\
        \hline
        More than one edge between two nodes & More packages than nodes \\
        \hline
        Edge to source class & Packages recursion \\
        \hline
        No edge from source class & Multiple connected components \\
        \hline
        Invalid node name & \multicolumn{1}{c}{\textbf{Unacceptable Defects}} \\
        \hline
        Invalid node id & Broken JSON \\
        \hline
        Invalid package id & Non-drawable diagram \\
        \hline
        Name of node not found in code & \\
        \hline
        Single-node package & \\
        \hline
        Class is in the package, but its method/field is not in the same package & \\
        \bottomrule
    \end{tabular}

    \caption{Defects list.}
\end{table}

\begin{table}[!t]
    \small    \centering\setlength{\tabcolsep}{2pt}

    \label{tab:defectsnum}
    \begin{tabular}{lcccccc}
    \toprule
       \textbf{Model} & \multicolumn{2}{c}{\textbf{Macro avg}} & \multicolumn{2}{c}{\textbf{Micro avg}} & \multicolumn{2}{c}{\textbf{Mean}} \\ 
       & \textbf{Low}& \textbf{Med} & \textbf{Low}& \textbf{Med} & \textbf{Low}& \textbf{Med} \\ \midrule
       {GPT-4o} & 0.299 & 0.074 & 0.290 & 0.069 & 1.479 & \textbf{0.354} \\ 
       {Claude 3.5}\newline {Sonnet} & 0.376 & 0.117 & 0.369 & 0.102 & 2.500 & 0.688 \\ \midrule
       \multicolumn{7}{c}{\textbf{Qwen2.5-Coder-14B}}\\\midrule
       {No SFT} & 0.283 & 0.283 & 0.249 & 0.081 & 2.833 & 0.917 \\
       {Claude Synth} & 0.218 & 0.181 & 0.277 & 0.112 & 2.770 & 1.125 \\ 
       {Fixed Claude Synth} & \textbf{0.144} & \textbf{0.043} & \textbf{0.162} & \textbf{0.038} & \textbf{1.583} & 0.375 \\ 
    \bottomrule
    \end{tabular}
    
    \caption{Aggregated number of defects (lower is better). Macro-Averaged: per-node then per-diagram. Micro-Averaged: overall per-node. Mean: per-diagram}
\end{table}

\begin{table}[!ht]
    \centering
    \label{tab:combined_metrics}
    \begin{tabular}{p{1.7cm}p{.9cm}|p{.75cm}p{.75cm}|p{.75cm}p{.75cm}p{.75cm}}
        \toprule
        \multicolumn{2}{c}{Metrics} & 
        \begin{sideways}GPT-4o\end{sideways} & 
        \begin{sideways}Claude-3.5-Sonnet\end{sideways} & 
        \begin{sideways}Qwen2.5-Coder-14B\end{sideways} & 
        \begin{sideways}\quad+SFT: Synth\end{sideways} & 
        \begin{sideways}\quad+SFT: Fixed Synth\end{sideways} \\ 
        
        \midrule
        
        \multirow{2}{*}{$\mathrm{\mathbf{Precision}}$}         & micro & 0.947& \textbf{0.954}& 0.754& 0.852& 0.859 \\
                                                               & macro & 0.943& \textbf{0.944}& 0.815& 0.863& 0.895\\ \hline
        \multirow{2}{*}{$\mathrm{\mathbf{Precision_{h}}}$}     & micro & 0.933& \textbf{0.940}& 0.653& 0.784 & 0.796\\
                                                               & macro & \textbf{0.940}& 0.938& 0.786& 0.828& 0.867\\\hline
        \multirow{2}{*}{$\mathrm{\mathbf{Recall}}$}            & micro & 0.439& 0.587& \textbf{0.778}& 0.773& 0.763\\
                                                               & macro & 0.540& 0.710& 0.797&  \textbf{0.805}& 0.799 \\\hline
        \multirow{2}{*}{$\mathrm{\mathbf{Recall_{h}}}$}        & micro & 0.580& 0.750& 0.808& \textbf{0.827}& \textbf{0.827}\\
                                                               & macro & 0.630& 0.836& 0.828& \textbf{0.845}& 0.842\\\hline
        \multirow{2}{*}{$\mathrm{\mathbf{F1}}$}                & micro & 0.600& 0.727& 0.766& \textbf{0.810}& 0.808\\
                                                               & macro & 0.648& 0.779& 0.765&  0.815& \textbf{0.823}\\\hline
        \multirow{2}{*}{$\mathrm{\mathbf{F1_{h}}}$}            & micro & 0.539& 0.668& 0.668& \textbf{0.730}& \textbf{0.730}\\  
                                                               & macro & 0.602& 0.742& 0.719&  0.763& \textbf{0.767} \\ 
        
        \bottomrule
    \end{tabular}
    \caption{Micro- and macro-averaged metrics ($\mathrm{\mathbf{h}}$ stands for ``hard''). {``Synth'' stands for \textit{Claude}-generated synthetic data.}}
\end{table}

\section{Evaluation results}\label{sec:evaluation}

\textbf{Models.}
We benchmark five models: \emph{GPT-4o}, \emph{Claude 3.5 Sonnet}, \emph{Qwen2.5-Coder-14B}, and two fine-tuned versions of \emph{Qwen2.5-Coder-14B} fine-tuned on diagram data (Section~\ref{sec:method}): trained on synthetic diagrams generated by \emph{Claude 3.5 Sonnet} ``as-is'' (Claude Synth) and after manual corrections (Fixed Claude Synth).

\textbf{Defect analysis.}
Table~\ref{tab:defectsnum} reveals that fine-tuning on high-quality, manually corrected data reduces diagram defects: the \emph{Qwen2.5-Coder-14B SFT: Fixed Claude Synth} model achieved the lowest defect counts, outperforming \emph{GPT-4o} and \emph{Claude 3.5 Sonnet}. Fine-tuning on raw synthetic data yielded ambiguous results. No unacceptable defects were observed in any experiment.

\textbf{Relevance evaluation.}
Two experts annotated \emph{48 diagrams per model} using the protocol from Section~\ref{sec:evaluationcriteria}. The inter-annotator agreement was high (Cohen's $\kappa = 0.82 \pm 0.02$), and all residual conflicts were resolved through discussion to produce a consensus gold set.
Tables~\ref{tab:labeling} and~\ref{tab:combined_metrics} show a pronounced precision–recall trade‑off among all base models. \emph{GPT‑4o} and \emph{Claude 3.5 Sonnet} deliver the highest precision, introducing almost no ``Hallucination'' or ``Verbosity'' nodes, but miss some relevant content, resulting in low recall. Conversely, the untuned \emph{Qwen2.5‑Coder‑14B} maximizes recall by producing many ``Sufficiency'' and ``Completeness'' nodes at the cost of the lowest precision.

The proposed fine-tuning scheme effectively closes this gap. The \emph{Qwen2.5-Coder-14B SFT: Fixed Claude Synth} (\emph{Qwen} tuned on manually corrected diagram data) achieves the best {F$_1$} scores, retaining the recall of the base \emph{Qwen} while markedly improving precision by suppressing ``Hallucinations'' and ``Verbosity'' nodes. Macro‑averaged metrics are consistently higher than micro‑averaged ones, indicating that all models fare better on smaller, less complex diagrams. The ``hard'' metrics reinforce the same picture: $\mathrm{Precision}_h$ is lower than standard precision for all models, while $\mathrm{Recall}_h$ is higher, implying that many true positives belong to the supplementary ``Completeness'' class and that models are most proficient at recovering indispensable ``Sufficiency'' nodes rather than non-essential ones.

Overall, our results demonstrate that fine‑tuning a code-specialized LLM even on a very small set of high‑quality, manually corrected diagrams produces diagrams that are both structurally sound and semantically faithful, making \emph{Qwen2.5-Coder-14B SFT: Fixed Claude Synth} the strongest candidate for practical applications in query‑driven code visualization.

\begin{figure}[!t]
    \centering
    \includegraphics[width=0.8\linewidth]{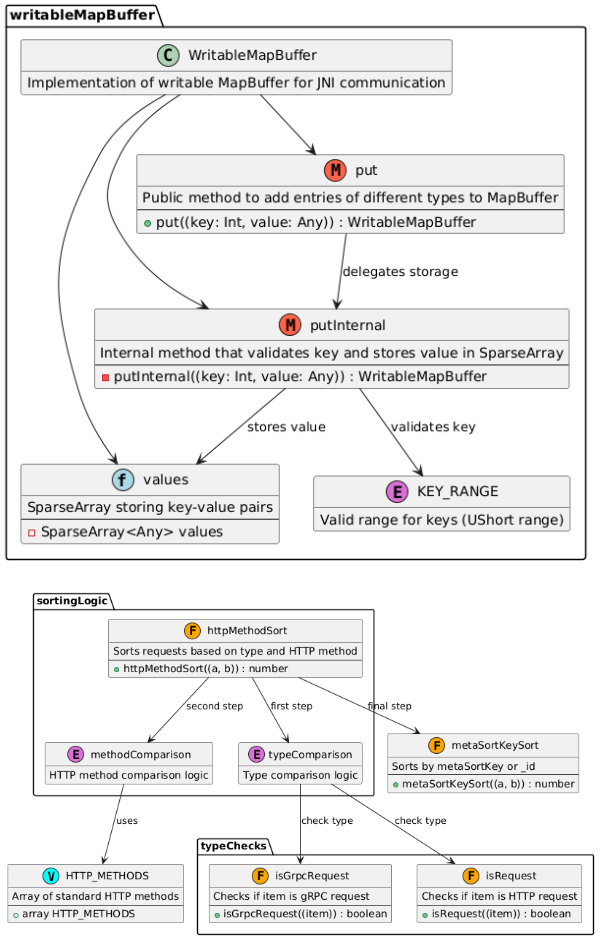}
    \caption{Sample generated diagrams.}
    \label{fig:diagelements}
\end{figure}

\textbf{Additional observations.}
We noted an intriguing strategy employed by models such as \emph{Claude 3.5 Sonnet}: generated nodes of type ``\textit{entity}'' (a UML node type) occasionally serve as abstractions of otherwise overly detailed structures (e.g., representing a database connection's parameters as several grouped entities rather than enumerating every individual parameter). This abstraction can facilitate comprehension of the response and reduce the cognitive load on the user by avoiding excessive detail. However, it complicates scoring because distinguishing legitimate abstractions from potential hallucinations remains challenging.

For qualitative evaluation, Fig.~\ref{fig:diagelements} shows two sample generated diagrams with all element types.

\section{Conclusion}\label{sec:conclusion}

In this work, we introduce query-driven UML diagram generation, a novel approach that bridges the gap between developer information needs and automated documentation. Our experiments have shown that LLMs can successfully generate focused, comprehensible diagrams that directly answer high-level queries about code architecture, design patterns, and component interactions. By fine-tuning Qwen2.5-Coder-14B on carefully curated data, we have achieved significant improvements in both structural correctness (lowest defect rates) and semantic relevance (best F1 scores), proving that even modest amounts of high-quality training data can effectively shape LLM behavior for specialized tasks.

Our dual evaluation framework---combining automatic defect analysis with human relevance assessment---provides a robust methodology for assessing diagram quality beyond simple syntactic correctness. The results validate both research questions: LLMs can generate relevant graph structures from code (\textbf{RQ1}), and targeted training successfully controls diagram quality properties (\textbf{RQ2}). The fine-tuned models show significant improvements in structural correctness with dramatically reduced defect rates, while maintaining or improving semantic relevance.

This study opens several promising research avenues for the future. First, extending beyond single-file analysis to generate diagrams that capture inter-module dependencies and project-wide architectural patterns (made practical with recent advances in long-context LLMs) would significantly increase practical utility. Second, conversational interfaces where developers can iteratively refine diagrams through follow-up queries (``zoom into the error handling'') would create better documentation experiences. Third, while we focused on class diagrams, the approach could extend to sequence diagrams (for execution flows), activity diagrams (for business logic), or state machines (for component lifecycles), each requiring specialized training data and evaluation metrics. Fourth, it would be great to explore hybrid approaches that combine LLM-generated diagrams with static analysis tools, looking for the best of both worlds. Finally, together with domain-specific adaptation, this could lead to the overall goal of \emph{real-time documentation}: by integrating query-driven generation into IDEs, developers could ultimately have on-demand, always-current documentation with zero user effort. This is the vision that we would like to fulfill in the future, and we hope that in this work we have taken important steps towards it.

\section*{Limitations and Future Directions}
While our results demonstrate the promise of query-driven diagram generation, several limitations point toward future improvements.

\emph{Training methodology}: our experiments focus on supervised fine-tuning (SFT). We explored alignment techniques, specifically ORPO~\cite{hong2024orpo}, using manually corrected diagrams as positive examples and their uncorrected counterparts as negatives, but obtained mixed results requiring further investigation. {A natural next step involves leveraging our automatic defect metrics as training signal in online reinforcement learning approaches.}

\emph{Evaluation scope}: although the natural language descriptions attached to diagram elements appear contextually appropriate and useful for understanding, we did not conduct human evaluation of their actual utility for developers. Similarly, descriptions of the relationships (edges) between nodes often seem plausible and meaningful, but high annotation cost prevented their systematic assessment.

\emph{Metric design}: our recall metrics estimate false negatives by comparing against the maximum relevant nodes found across all models for each query, providing upper bounds rather than absolute values. However, defining a single ``gold standard'' for diagram content remains theoretically challenging, as different developers may legitimately prefer different abstractions or levels of detail for the same query. To address this, our evaluation employs multiple safeguards (minimal/full versions, hard metrics, independent defect analysis) to ensure consistent model differentiation.

\emph{Context window}: our analysis is constrained to single code files, which limits practical applicability, as real-world developer queries often span multiple files. {However, our evaluation design—using large files aggregated from multiple repositories—naturally simulates challenges faced by retrieval-augmented generation (RAG) systems: irrelevant code (false positive retrievals) and incomplete external dependencies (false negative retrievals) mirror typical retrieval imperfections. Annotators were instructed to penalize both error types, allowing our results to demonstrate model robustness to imperfect inputs. Systematic evaluation with controlled FP/FN rates and integration with actual RAG pipelines remains future work.}

\emph{Generalization}: our experiments used a specific model family (Qwen2.5-Coder) and a relatively small training dataset, and broader validation would strengthen our claims.

We believe that these limitations, rather than diminishing our contributions, highlight the richness of this research direction. Each constraint above represents an opportunity for future work, moving towards intelligent documentation systems that adapt to developer needs.

\subsection*{Acknowledgments}
The work of A.~Alekseev was supported by the Ministry of Science and Higher Education of the Russian Federation (agreement 075-15-2025-344 dated 29/04/2025 for Saint Petersburg Leonhard Euler International Mathematical Institute at PDMI RAS).

\bibliographystyle{amsplain}
\bibliography{custom}

\end{document}